# Dynamics of Quantum Collapse in Energy Measurements

Ubaldo Tambini

*Dipartimento di Fisica, Università di Ferrara, and INFN, Sezione di Ferrara,*
*Via Paradiso 12, Ferrara, Italy 44100*

Carlo Presilla

*Dipartimento di Fisica, Università di Roma "La Sapienza", and INFN, Sezione di Roma,*
*Piazzale A. Moro 2, Roma, Italy 00185*

Roberto Onofrio

*Dipartimento di Fisica "G. Galilei", Università di Padova, and INFN, Sezione di Padova,*
*Via Marzolo 8, Padova, Italy 35131*

## Abstract

The influence of continuous measurements of energy with a finite accuracy is studied in various quantum systems through a restriction of the Feynman path-integrals around the measurement result. The method, which is equivalent to consider an effective Schrödinger equation with a non-Hermitian Hamiltonian, allows to study the dynamics of the wavefunction collapse. A numerical algorithm for solving the effective Schrödinger equation is developed and checked in the case of a harmonic oscillator. The situations, of physical interest, of a two-level system and of a metastable quantum-well are then discussed. In the first case the Zeno inhibition observed in quantum optics experiments is recovered and extended to nonresonant transitions, in the second one we propose to observe inhibition of spontaneous decay in mesoscopic heterostructures. In all the considered examples the effect of the continuous measurement of energy is a freezing of the evolution of the system proportional to the accuracy of the measurement itself.

03.65.Bz, 42.50.Wm, 79.60.Jv

Typeset using REVTEX

# I. INTRODUCTION

Technological progress in the measurement of physical quantities, expecially in quantum optics, in the physics of superconducting coherent devices such as SQUIDs, in the study of mesoscopic structures and in experimental gravitation, has revived the fifty-years old debates on foundations of quantum theory [1,2]. Collapse of the wavefunction, repeated measurements on a single object, inhibition of transitions under the effect of measurements are now also the language used by the experimentalists involved in this so-called quantum phenomenology. Ideal experiments which were thought in the thirties to sharp the paradoxical aspects of quantum theory when confronted to classical physics are now feasible. This originates a demand for quantitative predictions of quantum measurement theory. In general, among the observables of interest for the experimenters a crucial role is played by canonical coordinates and energy. The first kind of observable has been discussed extensively until now, expecially in connection to the fundamental limitations in the accuracy of position measurements [2,3]. In this paper we discuss a quantum measurement model which allows to describe the general features of an energy measurement independently of the particular measuring apparatus used to perform it. The model is based upon restriction of path-integrals around the measurement result. After a general description of the technique in Section II we analyze the case of a harmonic oscillator in Section III. This allows to check the numerical technique on a well established potential. The analysis turns out to more physical situations in Sections IV and V. In Section IV the case of a two-level system is analyzed and its link to the quantum Zeno effect is estabilished. In Section V the case of a properly designed heterostructure is analyzed to study the feasibility of an experiment exploiting spontaneous decay of localized electron states. This allows also for a comparison between the path-integral model of energy measurement and a more concrete one taking into account the detailed properties of a meter. In Section VI the conclusions are given: inhibition of the evolution of a system is obtained whenever measurements of energy are performed on it in the quantum regime of sensitivity.

# II. GENERAL FORMALISM

As a consequence of the interaction between the measured and the measuring systems, decoherentization arises. The phenomenon can be derived in the framework of standard quantum mechanics by summing over the degrees of freedom of the measuring system [4]. The measured system then evolves influenced by the interaction with the measuring system in such a way that the off-diagonal terms of its density matrix get damped. Of course, the exact evolution equation for the density matrix of the measured system depends on the specific nature of the measuring system, which is an inconvenient if general considerations are of interest. The path integral formalism with constraints added to force the system to move in a space region [5] or along an individual Feynman path [6] is a simple and effective tool for describing the essential features of a continuous measurement without reference to meter-dependent details. Here we will make use of the restricted path-integral approach proposed by Mensky [7,8].

Two different situations can arise: a prediction on the evolution of the measured system is requested before the continuous measurement is performed; a complete quantum mechanical description of the measured system, namely its wavefunction, is desired after the continuous measurement has been performed. In the first case, the *a priori analysis*, the outcome of the measurement is unknown and a sum over all possible results should be considered. The evolution of the measured system (during the measurement) is described by the density matrix. In the second case, the *a posteriori analysis*, the outcome of the measurement is known and a sum over all Feynman-paths compatible with the result should be considered. The measured-system evolution is described by the wavefunction. In both cases, the influence of the measurement on the evolution of the measured system is represented by a restriction of the possible choices (density matrix elements or paths) according to a weight functional. The strength of the restriction, measuring the amount of disturb per unit time fed by the measuring apparatus into the measured system, completely defines the measurement operation.

The *a posteriori analysis* has been shown to be very effective in understanding the accuracy of measurements of position in non-linear systems, monitored in a continuous [9] and impulsive way [10]. Here we will show how it applies to the case of measurements of energy. Let us suppose that a continuous measurement of energy with result $E(t)$ has been performed. The probability amplitude for the measured system going from point $q$ at time 0 to point $q'$ at time $t$ in the configuration space during the measurement is

$$K_{[E]}(q',t;q,0) = \int_q^{q'} d[q]d[p] e^{\frac{i}{\hbar}\int_0^t [p\dot{q} - H(p,q,t')]dt'} w_{[E]}[p,q]. \tag{1}$$

The weight functional $w_{[E]}$ restricts the integration around the measurement result. From a formal point of view a very simple and useful choice for $w_{[E]}$ is a Gaussian functional

$$w_{[E]}[q,p] = \exp\left(-\kappa \int_0^t (H(t') - E(t'))^2 dt'\right). \tag{2}$$

The constant $\kappa$ has dimensions $1/(\text{time} \times \text{energy}^2)$ and can be put in the form $\kappa^{-1} = \tau\epsilon^2$. The constants $\tau$ and $\epsilon$ have a well defined meaning in relation to the specific nature of the measuring apparatus. Konetchnyi, Mensky, and Namiot [11] have demonstrated that in the case of a continuous measurement of position the corresponding weight functional

$$w_{[x]}[q] = \exp\left(-\frac{1}{\tau\lambda^2}\int_0^t (q(t') - x(t'))^2 dt'\right) \tag{3}$$

coincides with that obtained for a measuring apparatus made by an infinite set of quantum oscillators, each interacting with the measured system within the length scale $\lambda$ and for a time interval $\tau$. A similar calculation in the case of a continuous measurement of energy allows us to interpret $\tau$ and $\epsilon$ as the characteristic time and energy scales of the interaction between the measuring and the measured systems. In other words $\tau$ and $\epsilon$ are the duration and the error of each microscopic measuring event. An interpretation in terms of the macroscopic measurement operation is also possible. If the measurement lasts a time $\Delta t = N\tau$ we can write $\kappa^{-1} = \tau\epsilon^2 = \Delta t \Delta E^2$ where $\Delta E = \epsilon/\sqrt{N}$. Here $\Delta E$ represents the error of the continuous measurement and it decreases, as expected, with the squared root of the measurement time.

Due to the choice of the weight functional, the probability amplitude can be written in the form

$$K_{[E]}(q',t;q,0) = \int_q^{q'} d[q]d[p] e^{\frac{i}{\hbar} \int_0^t [p\dot{q} - H_{eff}(p,q,t')]dt'} \tag{4}$$

where we introduced the effective Hamiltonian

$$H_{eff} = H - \frac{i\hbar}{\tau \epsilon^2}(H - E)^2. \tag{5}$$

Due to the presence of a fourth power of the momentum in $H_{eff}$ the functional integral (4) is meaningless from a rigorous point of view and we will use it only in a formal way to get a differential equation of motion. During the measurement the evolution of the system is given by the modified Schrödinger equation

$$i\hbar \frac{\partial}{\partial t} \psi(x,t) = H_{eff} \psi(x,t). \tag{6}$$

We note that the effective Hamiltonian (5) is not Hermitian. However, the eigenstates of the Hamiltonian $H$ of the unmeasured-system are also eigenstates of (5). We have just a modification of the time evolution of these eigenstates with a damping proportional to the difference between the corresponding eigenvalue and the measurement result. This allows us to dynamically describe the process of an incomplete and/or not instantaneous collapse of the state of the measured system.

Let us consider the case of $H$ with a discrete energy spectrum

$$H\phi_n(x) = E_n \phi_n(x). \tag{7}$$

Let $\psi(x,0)$ be the state of the system at the beginning of the measurement. We expand this state in the base $\{\phi_n\}$

$$\psi(x,0) = \sum_n c_n(0)\phi_n(x) \tag{8}$$

with the normalization $\sum_n |c_n(0)|^2 = 1$. By inserting (8) in (6) we have an evolution equation for the coefficients $c_n(t)$

$$\frac{d}{dt}c_n(t) = -\frac{i}{\hbar}\left[E_n - \frac{i\hbar}{\tau\epsilon^2}(E_n - E(t))^2\right]c_n(t) \tag{9}$$

with solution

$$c_n(t) = c_n(0)e^{-\frac{i}{\hbar}E_n t} e^{-\frac{1}{\tau\epsilon^2}\int_0^t (E_n - E(t'))^2 dt'}. \tag{10}$$

At the generic time $t$ during the measurement the measured system is completely described by the wavefunction

$$\psi(x,t) = \sum_n c_n(t)\phi_n(x). \tag{11}$$

Due to the presence of the anti-Hermitian part of $H_{eff}$, $\psi(x,t)$ loses its initial normalization according to the selection rule imposed by the measurement result. The probability to have the measured system in the eigenstate $n$ at time $t$ is

$$P_n(t) = \frac{|c_n(t)|^2}{\sum_m |c_m(t)|^2} = \frac{|c_n(0)|^2 e^{-\frac{2}{\tau \epsilon^2} \int_0^t (E_n - E(t'))^2 dt'}}{\sum_m |c_m(0)|^2 e^{-\frac{2}{\tau \epsilon^2} \int_0^t (E_m - E(t'))^2 dt'}}. \quad (12)$$

At the end of the measurement $t = \Delta t$, the probabilities $P_n$ have a simple interpretation in terms of the total error $\Delta E$ of the measurement. When $\Delta E \gg |E_n - E|$ (for simplicity we consider the case of a constant measurement result $E$ in the following discussion) the amount of back-action of the meter on the system is negligible and the probability to have the system in the eigenstate $n$ is close to that in absence of measurement. However, the above disequality can not be valid for all $n$ and a generic state of the system, i.e. containing contributions from all $n$, is always disturbed by the measurement at some extent. If the measurement result is some definite eigenvalue $E = E_i$, the wavefunction of the measured system approaches the eigenstate $\phi_i$ as much closely as much $\Delta E$ gets smaller. In the limit of an infinitely precise measurement a complete collapse of the wavefunction is obtained.

### III. MEASUREMENTS IN A HARMONIC OSCILLATOR

Here we apply the general formalism introduced in the previous section to the case of a harmonic oscillator described by the Hamiltonian

$$H = -\frac{\hbar^2}{2m} \frac{\partial^2}{\partial x^2} + \frac{1}{2} m \omega^2 x^2. \quad (13)$$

This simple system allows us to show with analytical tools the mechanism of the continuous collapse due to an energy measurement. It allows us also to test the numerical method described in appendix A that we will use in the analysis of more complicated systems.

The evolution of a wavefunction collapsing during an energy measurement is shown in Fig. 1. We consider an initial Gaussian state

$$\psi(x, 0) = \frac{1}{\sqrt{\sigma \sqrt{\pi}}} e^{-\frac{x^2}{2\sigma^2}} \quad (14)$$

which allows a simple analytical evaluation of $\psi(x,t)$ in terms of the harmonic oscillator eigenstates. The analytical results differ for less then 0.1 % from the numerical results shown in Fig. 1 obtained with the algorithm of Appendix A. In absence of measurement, i.e. $\Delta E = \infty$, the wavefunction remains a Gaussian with its width oscillating with period $\pi/\omega$. When the measurement is effective only the projection of $\psi(x,t)$ on the eigenstate with energy closest to the measurement result survives. The collapse is gradual and only for very precise and/or long lasting measurements it is fully accomplished. This behavior constitutes the basis of an effective relaxation method for computing numerically eigenvalues and eigenstates of Schrödinger operators [12].

Beside its simplicity, the harmonic oscillator is quite inappropriate for describing the interplay between energy measurements and stimulated transitions, a situation of wide interest in the applications, expecially in quantum optics. Indeed, due to the constant spacing of the spectrum, under a forcing term at resonance all the levels are occupied and no stationary regime for the stimulated transitions among the levels can be obtained. Systems with nonuniform level spacing or, for simplicity, two-level systems should be considered.

# IV. STIMULATED TRANSITIONS IN A TWO-LEVEL SYSTEM

The interplay between the transitions stimulated in a system by an external perturbation and the effect of a continuous measurement of energy have been recently discussed within the present formalism [13]. It turns out that the so-called quantum Zeno paradox introduced in [14] and [15,16] and observed in [17] is a particular example of the influence of the meter on the measured system (see also [18] for the debates following [17]). Here we discuss the differences between on-resonance and off-resonance perturbation. The case of off-resonance transitions is important expecially for schemes of measurements of energy in electromagnetic cavities based upon dispersive techniques recently proposed [19].

Let us return to the general formalism of Section II. Transitions among the levels of $H$ are obtained under the action of an appropriate external perturbation $V(t)$ which is added to the effective Hamiltonian of Eq. (5). The decomposition of the state $\psi(t)$ in terms of the eigenstates $\phi_n$ of the unmeasured and unperturbed system can be used again. The evolution equation for the coefficient $c_n$ contains also a term proportional to the perturbation strength and nondiagonal in the index $n$,

$$\frac{d}{dt}c_n(t) = -\frac{i}{\hbar}\left[E_n - \frac{i\hbar}{\Delta t \Delta E^2}(E_n - E(t))^2\right]c_n(t) - \frac{i}{\hbar}\sum_m V_{nm}(t)c_m(t) \qquad (15)$$

where $V_{nm}(t) = \langle \phi_n | V(t) | \phi_m \rangle$.

A particularly simple picture is obtained for a two-level system with energies $E_1$ and $E_2$. Assuming a perturbation potential $V_{11} = V_{22} = 0$ and $V_{12} = V_{21}^* = V_0 e^{i\omega(t-t_0)}$ with $V_0$ real, the solution of the system (15) is

$$c_1(t) = \exp\left\{-i\frac{E_1}{\hbar}t - \frac{(E_1-E)^2}{\Delta t \Delta E^2}t + iqt\right\}\left[c_1(0)\cos(wt) + \frac{qc_1(0) + e^{i\omega t_0}pc_2(0)}{iw}\sin(wt)\right] \qquad (16)$$

$$c_2(t) = \exp\left\{-i\frac{E_2}{\hbar}t - \frac{(E_2-E)^2}{\Delta t \Delta E^2}t - iqt\right\}\left[c_2(0)\cos(wt) - \frac{qc_2(0) - e^{-i\omega t_0}pc_1(0)}{iw}\sin(wt)\right] \qquad (17)$$

where $p = V_0/\hbar$, $q = \delta E/2\hbar + i\Omega$ with $\Omega = [(E_2 - E)^2 - (E_1 - E)^2]/2\Delta t \Delta E^2$ and $w = \sqrt{q^2 + p^2}$. The resonance condition is measured by the parameter

$$\delta E = \hbar\omega - (E_2 - E_1). \qquad (18)$$

In order to evidence the Zeno effect in a specific example let us suppose that initially the system is in the state 1 and the result of the continuous measurement is $E = E_1$. The probability $P_1(t)$ to have the system at time $t \leq \Delta t$ in the state 1 is

$$P_1(t) = \frac{|c_1(t)|^2}{|c_1(t)|^2 + |c_2(t)|^2} = \frac{1}{1 + \left|\frac{V_0}{\hbar\Omega - i\delta E/2 + \hbar w \cot(wt)}\right|^2} \qquad (19)$$

where

$$w = \frac{1}{\hbar}\sqrt{V_0^2 - \left(\hbar\Omega - \frac{i}{2}\delta E\right)^2} \qquad (20)$$

and

$$\Omega = \frac{(E_2 - E_1)^2}{2\Delta t \Delta E^2}. \qquad (21)$$

When the measurement error is large, i.e. $\Omega \to 0$, the system oscillates between levels 1 and 2 with Rabi frequency $\sqrt{4V_0^2 + \delta E^2}/\hbar$. Complete transitions to level 2, i.e. $P_1 = 0$, are obtained only with a resonant perturbation $\delta E = 0$. In the opposite limit of accurate measurements, when $w$ is imaginary, an overdamped regime is achieved in which transitions are inhibited. A critical damping is observed when $w$ is minimal, i.e. at a measurement error

$$\Delta E_{crit} = (E_2 - E_1)\sqrt{\frac{\hbar}{2V_0 \Delta t}} \qquad (22)$$

which defines the borderline between the Rabi-like behaviour ($\Delta E > \Delta E_{crit}$) and the Zeno-like inhibition ($\Delta E < \Delta E_{crit}$) in terms of the instrumental accuracy of the meter. The behavior of $P_1(t)$ is shown in Fig. 2 for $\delta E = 0$ and in Fig. 3 for $\delta E \neq 0$ for different choices of the measurement error $\Delta E$. In panels a) we show the Rabi-like behavior (solid line) in comparison with the corresponding results with no measurement performed (dashed line). Even in this regime the measurement has the effect to slightly decrease the transition frequency. In panels c) we show the Zeno-like behaviour (solid line) in comparison with the full inhibition occurring for $\Delta E \to 0$ (dashed line). In panels b) we show the behavior in the critical regime just above (solid line) and below (dot-dashed line) $\Delta E = \Delta E_{crit}$. Notice that in the off-resonance case the oscillations in the Rabi regime do not reach the zero even in absence of measurement. Moreover they get damped in presence of measurement even if the system is far from the critical region due to the imaginary term in Eq. (20). On the other hand no significant difference is observed between on-resonance and off-resonance cases in the Zeno regime. There the effect of the measurement is so strong to completely dominate any external perturbation.

## V. SPONTANEOUS DECAY FROM A QUANTUM WELL

The experiment designed to study quantum Zeno effect in atomic spectroscopy has been considered by some authors as not representative of the whole paradox because stimulated transition is used to connect two different levels. In this section we discuss a proposal for observation of quantum Zeno effect in systems subjected to spontaneous emission. The system we discuss is schematically depicted in the inset of Fig. 4. Metastable states are obtained in a semiconductor heterostructure where a barrier of width $L$ separates a well of width $l$ from a collector region. Let $f$ be the point separating the collector region $[f, +\infty[$ from the well region $]-\infty, f]$. For simplicity we suppose equal valence-band offsets $V_0$ and we neglect electron-electron interaction. Electrons generated inside the well, e.g. by a laser pulse, relax almost instantaneously in the $e1$ state (the low lying state of the quantum well) and then start tunneling outside the well. The validity of this picture implies a tunneling time much greater than the phonon relaxation time, i.e. a barrier width $L$ not too small. On the other hand we suppose the tunneling process faster than radiative and nonradiative electron-hole recombinations so that electrons leave the $e1$ of the well only via tunneling into the collector. In absence of measurements an exponential decay of the charge trapped in the well is obtained. As explained in Appendix B, the single-well metastable potential has a low lying resonance state $\phi_1(x)$ with complex eigenvalue $E_1 - i\Gamma_1/2$ which almost coincides in the well region with the $e1$ state $\phi_1^\infty(x)$. Electrons are initially in the $e1$ state, i.e. their wavefunction is $\psi(x,0) \simeq \phi_1(x)$ for $x \leq f$. The probability to have an electron in the region $]-\infty, f]$ at time $t$, the non-decaying probability, is

$$P(t) = \int_{-\infty}^{f} |\psi(x,t)|^2 dx \simeq \int_{-\infty}^{f} |\phi_1^\infty(x) e^{-\frac{i}{\hbar}(E_1 - i\Gamma_1/2)t}|^2 dx \simeq e^{-\Gamma_1 t/\hbar}. \quad (23)$$

Notice that $P(t)$ is related to the electronic charge $Q(t)$ measured in the collector region by $Q(t) = Ne[1 - P(t)]$, $N$ being the number of electrons photogenerated in the well and $e$ the electric charge.

Let us suppose that an energy measurement with known result is performed on the collected charge. A practical realization of such a kind of meter is obtained by growing a second barrier of width $L_m$ separated by a well of width $l_m$ from the measured system (see the inset of Fig. 5). The meter system acts as a filter allowing only the charge in a certain energy window to enter the collector region. Center and width of the energy window are specified by the well width $l_m$ and the barrier width $L_m$ of the meter, respectively. The energy measurement modifies the exponential decay of Eq. (23) depending on the measurement result and accuracy. We will report a comparison between the description in terms of the restricted path-integral method and the description in terms of the above-mentioned specific meter.

Since we are interested to follow in time the probability $P(t)$ (or the collected charge $Q(t)$) without specifying the duration of the measurement, it is convenient to express the measurement feedback through the microscopic quantities $\tau$ and $\epsilon$. Notice that there is only one degree of freedom in choosing these constants, the combination $\tau \epsilon^2$ only being relevant. We put $\tau = \hbar/\epsilon$ leaving $\epsilon$ as a free parameter. By assuming that the result of the energy measurement is constant $E = E_1$, the non-decaying probability is

$$P(t) \simeq \int_{-\infty}^{f} |\phi_1^\infty(x) e^{-\frac{i}{\hbar}(E_1 - i\Gamma_1/2)t} e^{-\frac{1}{\tau \epsilon^2}(E_1 - i\Gamma_1/2 - E)^2 t}|^2 dx \simeq e^{-\Gamma_{eff} t/\hbar} \quad (24)$$

where

$$\Gamma_{eff} = \Gamma_1 \left(1 - \frac{\Gamma_1}{2\epsilon}\right). \quad (25)$$

The decay of the electron measured to be inside the well in the $e1$ state is frozen proportionally with the measurement precision. Equation (25) is rigorously valid only for $\epsilon \gg \Gamma_1$. For $\epsilon < \Gamma_1/2$ it gives an absurd negative decay rate, however, already for $\epsilon \gtrsim \Gamma_1$ the results obtained from (25) are in good agreement with the numerical simulations reported in Fig. 4. In the numerical simulations $P(t)$ is obtained from the wavefunction evaluated in a space-time lattice with the algorithm described in Appendix A. We have chosen $l = 60$ Å, $L = 50$ Å, $V_0 = 0.1$ eV and an electron effective mass $m = 0.1\, m_0$, $m_0$ being the bare electron mass. With these parameters the first resonance level evaluated as explained in Appendix B is $E_1 = 3.72 \ 10^{-2}$ eV and $\Gamma_1 = 6.99 \ 10^{-4}$ eV.

Now we turn to the measurement performed with the model meter previously discussed. Since the measurement result has been chosen to be $E = E_1$, we have $l_m = l$. We put also $L_m = L$. The measured system couples quantum mechanically with the meter system in a well known manner. The previous first resonance splits into two resonant states $\phi_1(x)$ and $\phi_2(x)$ for which we evaluate $E_1 = 3.45 \ 10^{-2}$ eV, $\Gamma_1 = 2.86 \ 10^{-4}$ eV and $E_2 = 4.01 \ 10^{-2}$ eV, $\Gamma_2 = 4.35 \ 10^{-4}$ eV. Assuming that the electrons are initially in the $e1$ state of the unmeasured system, i.e. their wavefunction is $\psi(x,0) \simeq [\phi_1(x) + \phi_2(x)]/\sqrt{2}$, for $x \leq f$, we have

$$P(t) \simeq \frac{1}{2} \int_{-\infty}^{f} |\phi_1^{\infty}(x)e^{-\frac{i}{\hbar}(E_1 - i\Gamma_1/2)t} + \phi_2^{\infty}(x)e^{-\frac{i}{\hbar}(E_2 - i\Gamma_2/2)t}|^2 dx \simeq \frac{1}{2}\left(e^{-\Gamma_1 t/\hbar} + e^{-\Gamma_2 t/\hbar}\right). \quad (26)$$

Notice that the non-decaying probability includes integration over all the space but the collector region, i.e. $]-\infty, f]$ includes the meter. The rough prediction of Eq. (26) is shown in Fig. 5 (dashed line) in comparison with the corresponding exact numerical simulation (solid line). The exact probability $P(t)$ decays oscillating around the curve of Eq. (26) with frequency $(E_2 - E_1)/\hbar$. The model meter considered with barrier width $L_m = L$ is expected to represent an energy measurement of precision $\epsilon \sim \Gamma_1 \sim \Gamma_2$ and duration $\tau \sim \hbar/\Gamma_1$. Indeed, a comparison of Fig. 4 and Fig. 5 shows that this prediction is confirmed.

In the example we have discussed the realization of the back-action of the meter, i.e. its influence on the measured system, is quite evident. As the barrier acting as a meter measures the energy of the transmitted electrons with increasing accuracy, the resonances get narrower and the reflected current increases, coming back to the measured system. In the opposite limit, when the measurement is not accurate, the reflected current and the consequent back-action on the measured system are negligible.

The results obtained within the restricted path-integral formalism both in this section and in the previous one depend on the Gaussian ansatz (2) for the weight functional. This choice is not only mathematically convenient but corresponds to a specific, still quite general, model of measuring apparatus obeying the quantum-dynamical semigroup property [11,20]. Therefore qualitative agreement with the results obtained in a real measurement must be expected. On the other hand, detailed behaviors in a real measurement could be taken into account only by an accurate modellization of the employed meter. A comparison of Fig. 4 and Fig. 5 helps to clarify this point in a specific example. The restricted path-integral prediction correctly reproduces the main feature of the barrier-modelled meter, namely the variation of the decay time constant with respect to the unmeasured system case. On the other hand, the oscillations of $P(t)$ in Fig. 5 superimposed to the smooth decay curve (dashed line) have a period $\hbar\pi/(E_2 - E_1)$ and are the coherent oscillations of the charge between the well of the system and the well formed by the barrier meter. Of course these oscillations are not obtained in the restricted path-integral approach. This is a desirable feature when looking at a general formulation of quantum theory of measurement.

## VI. CONCLUSIONS

The time evolution of various systems during a continuous measurement of energy has been studied within a very general formalism. It has been shown that, regardless of the spe-cific realization of the meter, the measurement causes a partial freezing of the time evolution of the system. This fact is already known in quantum optics where a Zeno-like experiment has been performed. The generality of our approach allows not only to recover the quantum optics case but also to describe quantitatively a proposal for analogous experiments using man-made mesoscopic structures. In this case one could have the possibility to test Zeno inhibition on systems manifesting spontaneous decay with easier control of the parameters of the structure under measurement.

Our model can be considered as a quantitative dynamical approach to the energy-time uncertainty relationship already debated in the literature [21,22]. Let us define the influence time $\theta = \hbar/\Gamma_{eff} - \hbar/\Gamma_1$ expressing the modification of the spontaneous decay time under the influence of an energy measurement with error $\epsilon$. By using Eq. (25) in the appropriate limit of validity $\epsilon \gg \Gamma_1$ we have

$$\epsilon\,\theta \simeq \epsilon \left[ \frac{\hbar}{\Gamma_1}\left(1 + \frac{\Gamma_1}{2\epsilon}\right) - \frac{\hbar}{\Gamma_1} \right] \simeq \frac{\hbar}{2}. \tag{27}$$

High precision measurements of energy lead to changes of the evolution time-scales for the measured system. In the limit of an ideal, infinite accuracy, measurement the evolution of the system is completely frozen and the so called quantum Zeno effect is recovered.

## APPENDIX A: NUMERICAL SIMULATIONS

Here we briefly discuss the algorithm used for the numerical simulations of Eq. (6). The partial differential equation is transformed in a set of algebraic equations by discretizing the space-time $x - t$ in an appropriate two dimensional lattice. Let us suppose that we want to follow the temporal evolution of the initial state $\psi(x, 0)$ during the time interval $[0, t_{max}]$. Let $[x_{min}, x_{max}]$ be the space interval where the wavefunction $\psi(x, t)$ can be considered localized for $0 \leq t \leq t_{max}$. The space-time lattice is defined by putting

$$x \to x_{min} + j\Delta x \qquad j = 0, 1, 2, \ldots, J + 1 \tag{A1}$$

$$t \to n\Delta t \qquad n = 0, 1, 2, \ldots, N \tag{A2}$$

where the total number of points in the lattice is related to the lattice constants $\Delta x$ and $\Delta t$ by

$$(J + 1)\Delta x = x_{max} - x_{min} \tag{A3}$$

$$N\Delta t = t_{max}. \tag{A4}$$

The wavefunction reduces on the lattice to a matrix

$$\psi(x, t) \to \psi_j^n \tag{A5}$$

where $\psi_j^0$ are known. The effective Hamiltonian operator $H_{eff}$ is

$$H_{eff} = H - \frac{i\hbar}{\tau\epsilon^2}(H-E)^2 = \left(1 + \frac{2i\hbar}{\tau\epsilon^2}E\right)H - \frac{i\hbar}{\tau\epsilon^2}H^2 - \frac{i\hbar}{\tau\epsilon^2}E^2 \tag{A6}$$

and includes the action of the Hamiltonian $H = -(\hbar^2/2m)\partial^2/\partial x^2 + V(x,t)$ and its squared $H^2$ on $\psi(x,t)$. The action of $H$ on $\psi(x,t)$ gives

$$H\psi(x,t) \to (H\psi)_j^n = -\frac{\hbar^2}{2m\Delta x^2}(\psi_{j+1}^n - 2\psi_j^n + \psi_{j-1}^n) + V_j^n \psi_j^n \tag{A7}$$

where $V_j^n = V(x_j, t_n)$. For $H^2$ we have

$$H^2\psi(x,t) \to (H^2\psi)_j^n = \left(\frac{\hbar^2}{2m\Delta x^2}\right)^2 (\psi_{j+2}^n - 4\psi_{j+1}^n + 6\psi_j^n - 4\psi_{j-1}^n + \psi_{j-2}^n) + V_j^n V_j^n \psi_j^n$$
$$-\frac{\hbar^2}{2m\Delta x^2}\left[V_{j+1}^n \psi_{j+1}^n - 2V_j^n \psi_j^n + V_{j-1}^n \psi_{j-1}^n + V_j^n(\psi_{j+1}^n - 2\psi_j^n + \psi_{j-1}^n)\right]. \tag{A8}$$

The discretization of the operator $i\hbar\partial/\partial t$ is obtained with the Cayley approximation for the exponential time evolution operator

$$e^{-\frac{i}{\hbar}K\Delta t} \simeq \left[1 + \frac{i\Delta t}{2\hbar}K\right]^{-1}\left[1 - \frac{i\Delta t}{2\hbar}K\right] \tag{A9}$$

which preserves unitarity for $K$ Hermitian and is correct $\mathcal{O}(\Delta t^2)$ for $K$ time-independent. In our case $H_{eff}$ depends on time and is the sum of a Hermitian term and an anti-Hermitian one. The Cayley approximation is still the right discretization scheme since it avoids the stability problems arising from the Hermitian part. The finite difference equation for the time evolution is the Crank-Nicholson scheme

$$\left[1 + \frac{i\Delta t}{2\hbar}H_{eff}(t+\Delta t)\right]\psi(t+\Delta t) = \left[1 - \frac{i\Delta t}{2\hbar}H_{eff}(t)\right]\psi(t) \tag{A10}$$

which corresponds in the lattice to the following linear system

$$\gamma\psi_{j+2}^{n+1} + \beta_j^{n+1}\psi_{j+1}^{n+1} + (1+\alpha_j^{n+1})\psi_j^{n+1} + \beta_{j-1}^{n+1}\psi_{j-1}^{n+1} + \gamma\psi_{j-2}^{n+1} =$$
$$-\gamma\psi_{j+2}^n - \beta_j^n\psi_{j+1}^n + (1-\alpha_j^n)\psi_j^n - \beta_{j-1}^n\psi_{j-1}^n - \gamma\psi_{j-2}^n \tag{A11}$$

where

$$\gamma = \frac{i\Delta t}{2\tau\epsilon^2}\left(\frac{\hbar^2}{2m\Delta x^2}\right)^2 \tag{A12}$$

$$\beta_j^n = -4\gamma - \frac{i\hbar\Delta t}{2m\Delta x^2}\left[\frac{1}{2} + \frac{i\hbar}{2\tau\epsilon^2}(2E - V_{j+1}^n - V_j^n)\right] \tag{A13}$$

$$\alpha_j^n = 6\gamma + \frac{i\hbar\Delta t}{2m\Delta x^2}\left[1 + \frac{2i\hbar}{\tau\epsilon^2}(E - V_j^n)\right] + \frac{i\Delta t}{2\hbar}\left[V_j^n - \frac{i\hbar}{\tau\epsilon^2}(E - V_j^n)^2\right]. \tag{A14}$$

By using the condition $\psi_{-1}^n = \psi_0^n = \psi_{J+1}^n = \psi_{J+2}^n = 0$ which reflects the fact that the wavefunction is localized in $[x_{min}, x_{max}]$, we can write Eq. (A11) in matrix notation

$$\left(\delta_{ij} + \mathcal{R}_{ij}^{n+1}\right) \psi_j^{n+1} = \left(\delta_{ij} - \mathcal{R}_{ij}^{n}\right) \psi_j^{n} \qquad i,j = 1,\ldots,J \tag{A15}$$

where the matrix $\mathcal{R}^n$ is 5-diagonal

$$\mathcal{R}^n = \begin{pmatrix} \alpha_1^n & \beta_1^n & \gamma & 0 & \ldots & 0 & 0 & 0 & 0 \\ \beta_1^n & \alpha_2^n & \beta_2^n & \gamma & \ldots & 0 & 0 & 0 & 0 \\ \gamma & \beta_2^n & \alpha_3^n & \beta_3^n & \ldots & 0 & 0 & 0 & 0 \\ 0 & \gamma & \beta_3^n & \alpha_4^n & \ldots & 0 & 0 & 0 & 0 \\ \vdots & \vdots & \vdots & \vdots & \ddots & \vdots & \vdots & \vdots & \vdots \\ 0 & 0 & 0 & 0 & \ldots & \alpha_{J-3}^n & \beta_{J-3}^n & \gamma & 0 \\ 0 & 0 & 0 & 0 & \ldots & \beta_{J-3}^n & \alpha_{J-2}^n & \beta_{J-2}^n & \gamma \\ 0 & 0 & 0 & 0 & \ldots & \gamma & \beta_{J-2}^n & \alpha_{J-1}^n & \beta_{J-1}^n \\ 0 & 0 & 0 & 0 & \ldots & 0 & \gamma & \beta_{J-1}^n & \alpha_J^n \end{pmatrix}. \tag{A16}$$

Starting from $\psi_j^0$ the solution of the above system at each time step gives the unknown $\psi_j^{n+1}$ in terms of the known $\psi_j^n$. Notice that the matrix $\mathcal{R}^n$ does not depend on time if the potential $V$ is time independent.

Due to its 5-diagonal nature the linear system in Eq. (A15) is solved by LU decomposition accomplished through standard-library subroutines. The number of iterations required for the solution grows linearly with the matrix dimension $J$. The computer time needed for the simulation of the full evolution is about $CNJ$, with $C \simeq 7.6$ $\mu$s in a VAX 7000-610 machine.

## APPENDIX B: RESONANCE STATES OF THE METASTABLE POTENTIAL

We want to derive a semiclassical formula for the resonances $\lambda_n = E_n - i\Gamma_n/2$ of the metastable square-well potential

$$V(x) = V_0 \, 1_{[-\infty,a]} + V_0 \, 1_{[b,c]} + V_0 \, 1_{[d,f]} \tag{B1}$$

where $a < b < c < d < f$ and $1_{[a,b]}$ means 1 for $a < x < b$ and 0 elsewhere. The eigenfunctions of the stationary Schrödinger equation

$$\left[-\frac{\hbar^2}{2m}\frac{\partial^2}{\partial x^2} + V(x)\right]\phi_n(x) = \lambda_n \phi_n(x) \tag{B2}$$

are such that $\phi_n(x) \propto \exp\left(i\sqrt{\lambda_n}x\right)$, for $x \to \infty$. We also observe that, as a special case of the method of complex scaling [23], $\phi_n(x)$ is of class $L^2$ on the contour $\gamma = ]-\infty, f] \cup e^{i\theta} [f, +\infty[$ if $\theta > 0$ is conveniently chosen. In the limit of $f - d$ large, we approximate $E_n$ with the eigenvalues of the confining potential $V^\infty(x) = V_0 \, 1_{[-\infty,a]} + V_0 \, 1_{[b,c]} + V_0 \, 1_{[d,+\infty[}$. The functions

$$\phi_n^\infty(x) = A \begin{cases} e^{k_0 x} & x < a \\ B^+ e^{ikx} + B^- e^{-ikx} & a < x < b \\ C^+ e^{k_0 x} + C^- e^{-k_0 x} & b < x < c \\ D^+ e^{ikx} + D^- e^{-ikx} & c < x < d \\ F^+ e^{k_0 x} + F^- e^{-k_0 x} & d < x \end{cases} \tag{B3}$$

where

$$k = \sqrt{\frac{2m}{\hbar^2}E_n} \tag{B4}$$

and

$$k_0 = \sqrt{\frac{2m}{\hbar^2}(V_0 - E_n)} \tag{B5}$$

are the bound states for the potential $V^\infty(x)$ when $F^+ = 0$. The $C^1$ requirement at the points $a$, $b$, $c$ and $d$ gives:

$$B^\pm = \frac{e^{\mp ika}}{2ik} e^{k_0 a}(ik \pm k_0) \tag{B6}$$

$$C^\pm = \frac{e^{\mp k_0 b}}{2k_0}\left[B^+ e^{ikb}(k_0 \pm ik) + B^- e^{-ikb}(k_0 \mp ik)\right] \tag{B7}$$

$$D^\pm = \frac{e^{\mp ikc}}{2ik}\left[C^+ e^{k_0 c}(ik \pm k_0) + C^- e^{-k_0 c}(ik \mp k_0)\right] \tag{B8}$$

$$F^\pm = \frac{e^{\mp k_0 d}}{2k_0}\left[D^+ e^{ikd}(k_0 \pm ik) + D^- e^{-ikd}(k_0 \mp ik)\right] \tag{B9}$$

The eigenvalues $E_n$ are found by solving numerically the equation $F^+(E_n) = 0$. The normalization of $\phi_n^\infty(x)$ is achieved by choosing

$$A^{-2} = \frac{e^{2k_0 a}}{2k_0} + \left(|B^+|^2 + |B^-|^2\right)(b-a) + 2\Re e\left\{B^+ B^{-*} e^{ik(a+b)}\right\}\frac{\sin[k(b-a)]}{k}$$

$$+2\Re e\left\{C^+ C^{-*}\right\}(c-b) + |C^+|^2\frac{e^{2k_0 c} - e^{2k_0 b}}{2k_0} + |C^-|^2\frac{e^{-2k_0 b} - e^{-2k_0 c}}{2k_0}$$

$$+\left(|D^+|^2 + |D^-|^2\right)(d-c) + 2\Re e\left\{D^+ D^{-*} e^{ik(c+d)}\right\}\frac{\sin[k(d-c)]}{k} + |F^-|^2\frac{e^{-2k_0 d}}{2k_0}. \tag{B10}$$

Once $E_n$ is known, the corresponding $\Gamma_n$ can be found by observing that in the limit of $f - d$ large we have the semiclassical approximation

$$\phi_n(x) = A\begin{cases} e^{k_0 x} & x < a \\ B^+ e^{ikx} + B^- e^{-ikx} & a < x < b \\ C^+ e^{k_0 x} + C^- e^{-k_0 x} & b < x < c \\ D^+ e^{ikx} + D^- e^{-ikx} & c < x < d \\ F^+ e^{k_0 x} + F^- e^{-k_0 x} & d < x < f \\ G^+ e^{ikx} & f < x \end{cases} \tag{B11}$$

with $A$, $B^\pm$, $C^\pm$, $D^\pm$ and $F^-$ evaluated as above and $F^+$, $G^+$ obtained through the $C^1$ requirement at $f$

$$F^+ = F^- \frac{k_0 + ik}{k_0 - ik} e^{-2k_0 f} \tag{B12}$$

$$G^+ = F^- \frac{2k_0}{k_0 - ik} e^{-k_0 f - ikf}. \tag{B13}$$

By multiplying the stationary Schrödinger equation (B2) with $\phi_n^*$ and the complex conjugate equation with $\phi_n$ and subtracting the two results, after integration in $]-\infty, f^+]$, we find

$$\Gamma_n \int_{-\infty}^{f^+} |\phi_n(x)|^2 dx = \frac{\hbar^2}{2m} 2\Im m \left\{ \frac{d\phi_n(f^+)}{dx} \phi_n(f^+)^* \right\}. \tag{B14}$$

By using Eq. (B13) in the r.h.s. and approximating $\phi_n \simeq \phi_n^\infty$ in the integral, we get:

$$\Gamma_n = \frac{\hbar^2}{2m} \frac{8kk_0^2}{k^2 + k_0^2} \frac{|AF^- e^{-k_0 f}|^2}{1 - |AF^- e^{-k_0 f}|^2 / 2k_0}. \tag{B15}$$

The above results apply also to the case of a single-well metastable potential by putting $a = b = c$.

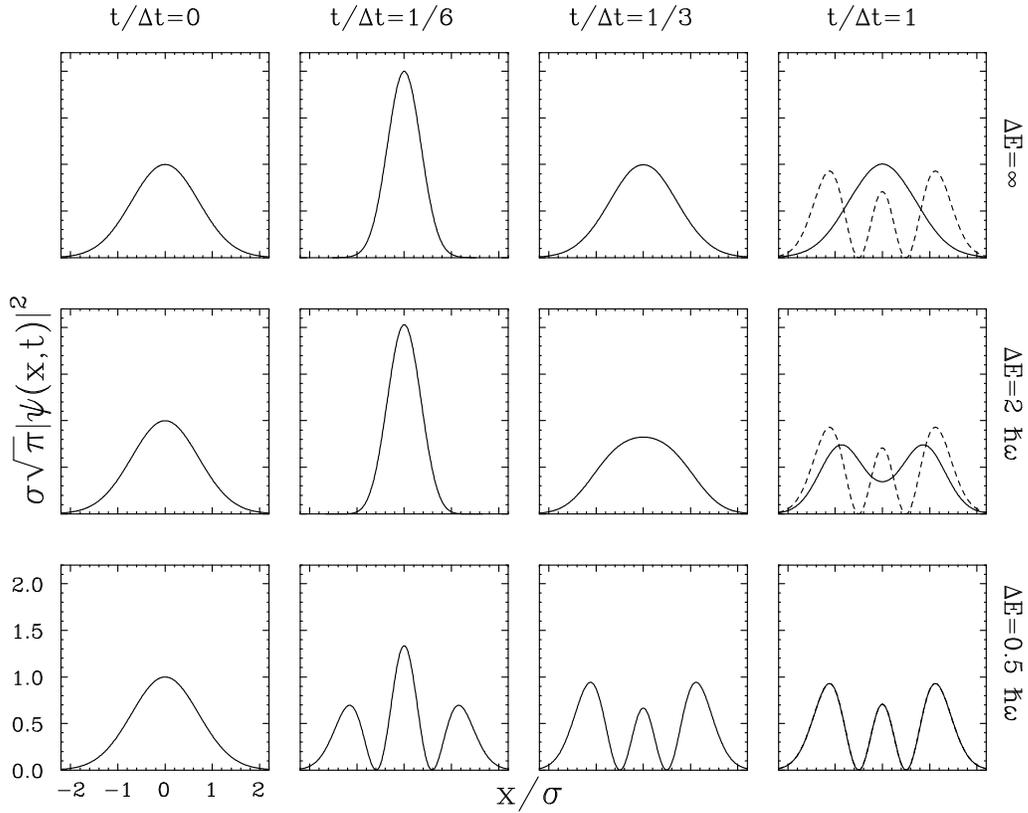

FIG. 1. Evolution of the wavefunction collapse under a continuous measurement of energy in a harmonic oscillator. Each panel shows, at different times and for different measurement errors, the squared modulus of the wavefunction during a measurement with constant result $E = E_2 = \frac{5}{2}\hbar\omega$. The initial wavefunction is a Gaussian of width $\sigma = \sqrt{2\hbar/m\omega}$ and we put $\Delta t = 3\pi/\omega$, $\hbar = 2m = 1$, $\omega = 5 \cdot 10^{-4}$. In the right-most panels corresponding to the time $t = \Delta t$ the dashed line indicates the squared modulus of the eigenfunction $\phi_2$ towards which the initial state collapses for a finite $\Delta E$.

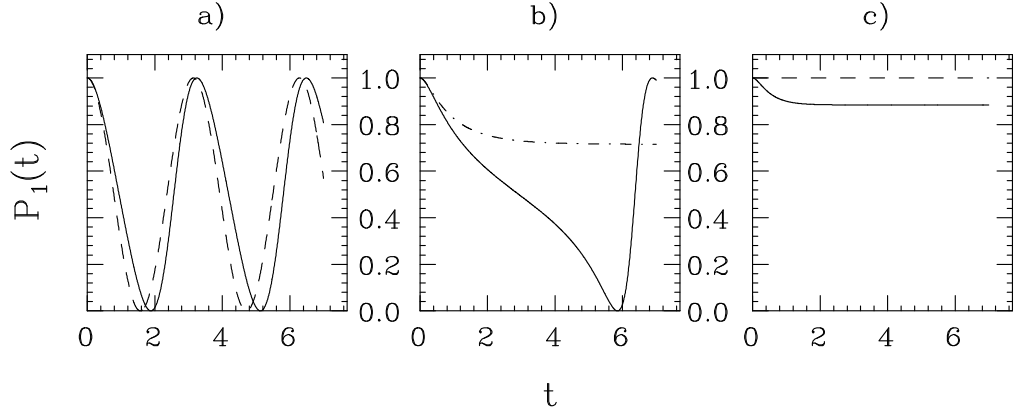

FIG. 2. Probability $P_1(t)$ to observe a two-level system in the state 1 during a measurement of energy and under the influence of a resonant perturbation. The system is in the state 1 at time $t = 0$ and the result of the measurement is $E = E_1$ constant. The measurement error is $\Delta E = \infty$ (dashed line) and $\Delta E = 2\Delta E_{crit}$ (solid line) in panel a), $\Delta E = 1.07\Delta E_{crit}$ (solid line) and $\Delta E = 0.99\Delta E_{crit}$ (dot-dashed line) in panel b) and $\Delta E = 0.5\Delta E_{crit}$ (solid line) and $\Delta E = 0$ (dashed line) in panel c). We put $\Delta t = 10\pi\hbar/V_0$ and $E_2 - E_1 = V_0 = \hbar = 1$.

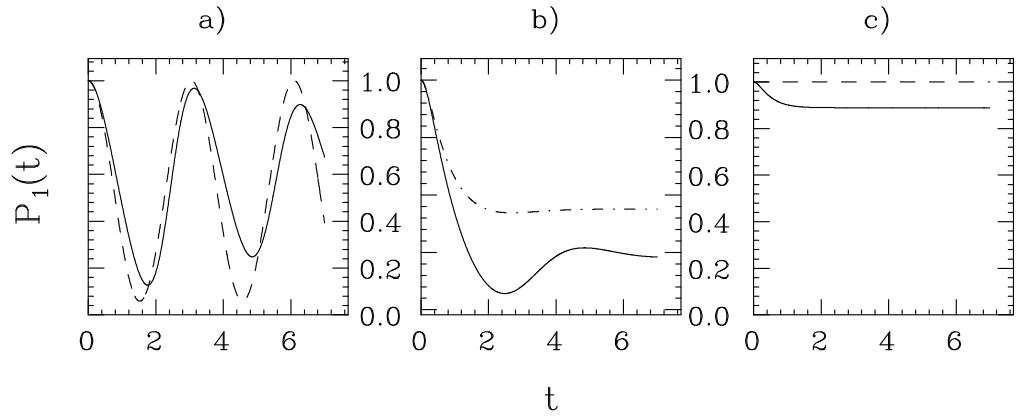

FIG. 3. As in Fig. 2 but for a nonresonant perturbation with $\delta E = (E_2 - E_1)/2$.

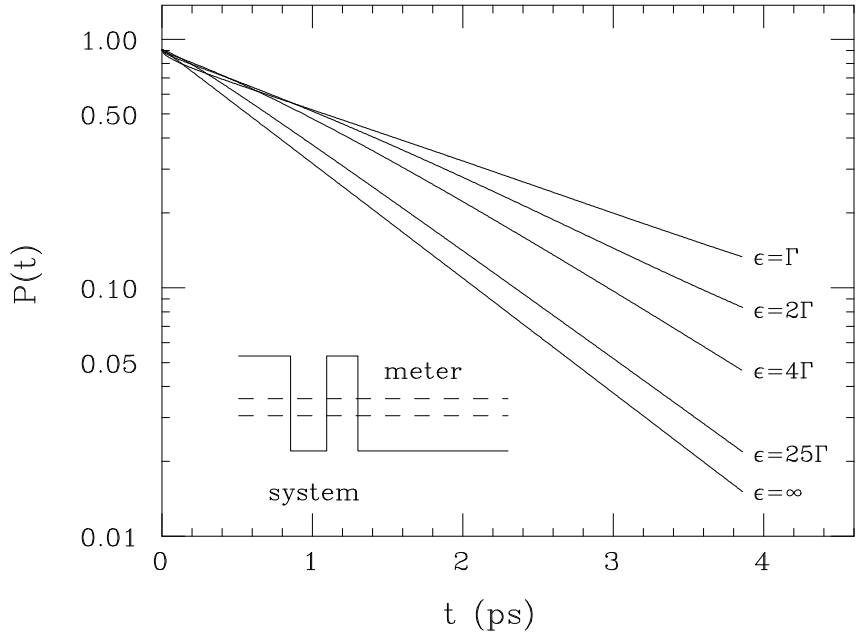

FIG. 4. Non-decaying probability $P(t)$ for electrons generated inside the metastable well shown in the inset during a continuous measurement of energy achieved by restriction of the Feynman paths. The measurement result is the energy of the resonant state and we put $\tau = \hbar/\epsilon$ for different values of $\epsilon$.

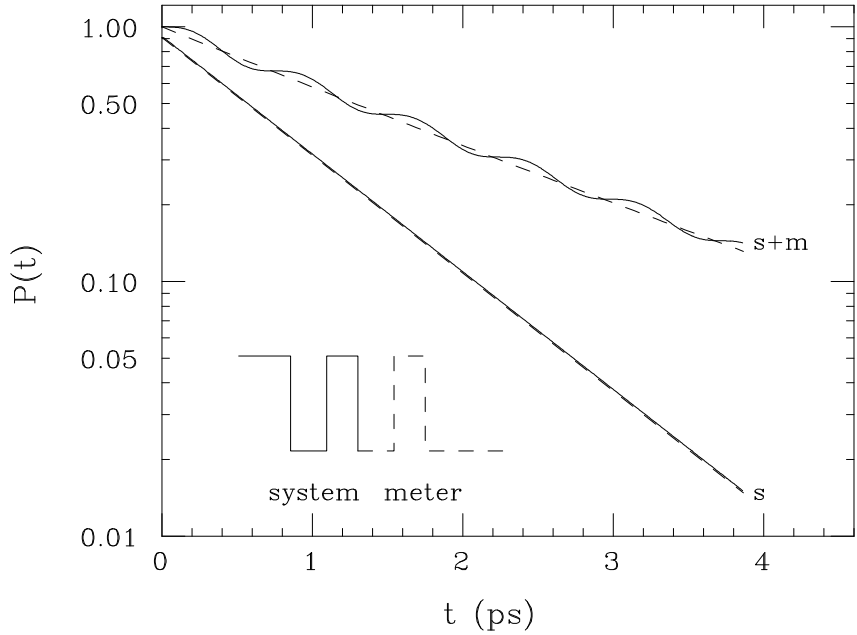

FIG. 5. Non-decaying probability $P(t)$ (upper solid line) for electrons generated inside the same metastable well of Fig. 4 but during a continuous measurement of energy achieved by the model meter shown in the inset (dashed line). The lower solid line is obtained in absence of measurement. The dashed lines are the results of Eq. (23) (lower line) and Eq. (26) (upper line).